\chapter{Black holes \label{C5}}

\section{Horizons \label{S5.1}}

The example of gravitational shock waves/domain walls has shown that 
space-time in four dimensions is a truely dynamical arena for physics
even in the absence of matter, although adding matter and other fields
of force ---for example in the form of scalar or electromagnetic 
fields--- makes the possible structures even richer. In this 
chapter we turn to another class of very interesting structures 
in four-dimensional space-time: static or quasi-static solutions 
of the Einstein equations which represent gravitating extended bodies, 
whose static fields become strong enough to capture permanently 
anything that gets sufficiently close to the central core. As these 
quasi-static objects even capture light they are called black holes, 
but it has turned out that their properties are much more interesting 
and peculiar than this ominous but dull sounding name suggests. 

Black holes possess two characteristic features. The first, which has
given them their name, is the existence of a horizon: a closed surface 
through which objects can fall without ever being able to return. The 
second is the presence of a singularity, a locus of points inside the 
horizon where the curvature becomes infinite. It has even been 
conjectured by some authors that these two properties are so closely 
linked that any space-time singularity should always be hidden behind 
a horizon. This conjecture bears the name of the Cosmic Censorship 
hypothesis. Although as an unqualified mathematical theorem the 
hypothesis is certainly not correct, it could be true for singularities 
that arise in realistic processes from the gravitational collapse of 
macroscopic bodies, like imploding massive stars. 

In this chapter we restrict our attention to the special black-hole 
solutions that might be called classical `eternal' black holes, in 
the sense that at least outside the horizon they describe stationary 
gravitational fields, and do not require for the description of their
behaviour the inclusion of their formation from collapsing matter. 
In fact they are solutions of the source-free Einstein or coupled 
Einstein-Maxwell equations and are characterised for an outside 
observer completely in terms of their mass, their angular momentum and 
their electro-magnetic charge. One might alternatively think of them
as black holes that have been formed in the infinite past and settled
in a stationary state in which all trace of their previous history has
been lost, e.g.\ by emitting gravitational radiation.

Because of this restriction these lectures do not include a detailed 
account of the process of gravitational collapse as studied by 
astrophysicists, but they do allow one to isolate and study in detail 
the special aspects of the physics of black-holes. It may even be that 
these very special simple solutions are relevant to elementary particle 
physics near the Planck scale, but that leads one to consider the 
problems of quantum gravity, a subject we will not deal with at length 
in these lectures. The interested reader can find more information on
many aspects of black-hole structure and formation in the literature
\ct{Chan,Nov}. 

\section{The Schwarzschild solution \label{S5.2}}

The simplest of all stationary black-hole solutions of the source-free Einstein
equations is that for the static spherically symmetric space-time, 
asymptotically flat at spatial infinity, first described in the literature by 
Schwarzschild\footnote{The solution was found independently at the same time 
in \ct{Dros}.} \ct{Schw}. We present a derivation of this solution making full 
use of the symmetries and their connection to Killing vectors explained in 
chapter \ref{C2}. 

In view of the spherical symmetry we introduce polar coordinates 
$(r,\th,\vf)$ in three-dimensional space and add a time coordinate $t$,
measuring asymptotic Minkowski time at $r \rightarrow \infty$. The 
Schwarzschild solution is obtained by requiring that there exists
at least one coordinate system parametrized like this in which the 
following two conditions hold: \nl
$(i)$ the metric components are $t$-independent; \nl
$(ii)$ the line-element is invariant under the three-dimensional rotation
group $SO(3)$ acting on three-vectors {\boldmath{$r$}} in the standard
(linear) way. \nl 
The first requirement is of course equivalent to invariance of the 
metric under time-shifts $t \rightarrow t + \del t$. Thus formulated both 
conditions take the form of a symmetry requirement.

In chapter (\ref{C2}) we discussed how symmetries of the metric are 
related to Killing vectors. Rotations of three-vectors are generated by 
the differential operators {\boldmath{$L$}} of orbital angular momentum:

\be
\ba{ll}
L_1\, = \, \dsp{ 
          -\sin \vf\, \dd{}{\th}\, -\, \cot \th \cos \vf\, \dd{}{\vf}, } & 
L_2\, =\, \dsp{ 
          \cos \vf\, \dd{}{\th}\, -\, \cot \th \sin \vf\, \dd{}{\vf}, }\\
 & \\
L_3\, =\, \dsp{ \dd{}{\vf}. }  &     
\ea
\label{5.1}
\ee

\nit
Similarly, time-shifts are generated by the operator

\be
L_0\, =\, \dd{}{t}.
\label{5.2}
\ee

\nit
According to the discussion in sect.(\ref{S2.4}) the coefficients of the 
symmetry operators $L_A = L_A^{\:\:\mu} \pl_\mu$ are to be components 
of a set of Killing vectors $L_A$, $A = 0,1,2,3$:

\be
D_\mu L_{A \nu}\, +\, D_{\nu} L_{A \mu}\, =\, 0.
\label{5.2.1}
\ee 

\nit
These can equivalently be characterised as a set of constants of geodesic 
flow as in eq.(\ref{c2.42}):

\be
J_{A}\, =\,  L_A^{\:\:\mu} p_\mu. 
\label{5.3}
\ee

\nit
The relation with the metric is, that the Poisson brackets of these 
constants of geodesic flow with the hamiltonian are to vanish:

\be
\lacc J_{A}, H \racc\, =\, 0, \hspace{3em}
H\, =\, \frac{1}{2}\, g^{\mu\nu} p_\mu p_\nu.
\label{5.4}
\ee

\nit
The general solution of these conditions for the hamiltonian $H$ is 
$t$-independent, whilst it can depend on the angular coordinates only 
through the Casimir invariant 

\be 
\bL^2\, =\, p_\th^2\, +\, \frac{p_\vf^2}{\sin^2 \th}.
\label{5.4.1}
\ee

\nit
It follows that the metric must be of the form\footnote{In this chapter
we employ natural units with $c = 1$.}

\be
ds^2\, =\, - h^2(r) dt^2\, +\, g^2(r) dr^2\, +\, k^2(r) d\Og^2,
\label{5.5}
\ee

\nit
where $d\Og^2 = d\th^2 + \sin^2 \th\, d\vf^2$ is the angular distance 
on the unit sphere, and the coefficients $h(r)$, $g(r)$ and $k(r))$ are 
functions of the radial coordinate $r$ only. Indeed, the hamiltonian
derived from this metric is

\be
\ba{lll} 
H & = & \dsp{ -\, \frac{1}{2h^2}\, p_t^2\, +\, \frac{1}{2g^2}\, p_r^2\,
  +\, \frac{1}{2k^2}\, \lh p_\th^2 + \frac{1}{\sin^2 \th}\, p_\vf^2
  \rh  }\\ 
 & & \\
 & = & \dsp{ -\, \frac{1}{2h^2}\, p_t^2\, +\, \frac{1}{2g^2}\, p_r^2\,
        +\, \frac{\bL^2}{2k^2}.  }
\ea
\label{5.5.1}
\ee

\nit
Note that the set of functions $(h,g,k)$ can not be unique, as one can 
always perform a coordinate transformation $r \rightarrow \bar{r}(r)$ to 
another metric of the same general form (\ref{5.5}) with coefficients 
$(\bar{h},\bar{g},\bar{k})$. There are several standard options to 
remove this freedom; to begin with, we choose one in which the spherical 
symmetry is manifest and the spatial geometry is as close to that of 
flat space as possible. That is, we take $r$ as the solution of 

\be
k^2(r)\, =\, r^2 g^2(r).
\label{5.6}
\ee

\nit
Then the line element becomes

\be
ds^2\, =\, - h^2(r) dt^2\, +\, g^2(r) \lh dr^2\, +\, r^2 d\Og^2 \rh\,
       =\, - h^2(r) dt^2\, +\, g^2(r) d\br^{\, 2}. 
\label{5.7}
\ee

\nit
Clearly it is rotation invariant, and the three-dimensional space-like 
part of the line element is conformal to that of flat space in polar 
coordinates, with the conformal factor $g(r)$ depending only on the radius, 
and not on the orientation or the time. In the literature the coordinates 
chosen here to parametrize the Schwarzschild metric are called isotropic 
coordinates. 

Up to the freedom of radial reparametrizations, the form of the line 
element (\ref{5.7}) is completely determined by the symmetry requirements. 
To fix the radial dependence of the remaining coefficients $h(r)$ and 
$g(r)$ we finally substitute the metric into the Einstein equations. For 
the Einstein tensor $G_{\mu\nu} = R_{\mu\nu} - 1/2\, g_{\mu\nu} R$ we 
find that it is diagonal with components 

\be
\ba{lll}
G_{tt} & = & \dsp{ \frac{h}{g^3}\, \lh 2h (g^{\prime\prime} + \frac{2}{r} 
  g^{\prime}) - \frac{hg^{\prime\, 2}}{g} \rh, }\\
 & & \\
G_{rr} & = & \dsp{ - \frac{1}{gh}\, \lh 2 h^{\prime} g^{\prime} + 
  \frac{hg^{\prime\, 2}}{g} + \frac{2}{r}\, (hg)^{\prime} \rh, }\\
 & & \\ 
G_{\th\th} & = & \dsp{ \frac{G_{\vf\vf}}{\sin^2 \th}\, =\,
  - \frac{r^2}{gh}\, \lh h^{\prime\prime} g + h g^{\prime\prime} + 
  \frac{1}{r}\, (hg)^{\prime} - \frac{hg^{\prime\, 2}}{g} \rh, }
\ea
\label{5.8}
\ee

\nit 
In the absence of sources or a cosmological constant these expressions
have to vanish. As boundary conditions at infinity, where space-time must  
become asymptotically flat, we require that $r \rightarrow \infty$ implies 
$h \rightarrow 1$ and $g \rightarrow 1$. Then we have three differential 
equations for two unknown functions, and therefore the equations must be 
degenerate. One can check, that this is the case if the following two
relations are satisfied:

\be
hg\, =\, g\, +\, r\pr{g}, \hspace{3em} 
  \left[ r^3 (hg)^{\prime} \right]^{\prime}\, =\, 0.
\label{5.9.1}
\ee

\nit
Indeed, in this case the various components of $G_{\mu\nu}$ become
proportional:

\be
G_{rr}\, = \, - \frac{g^2}{h^2}\, G_{tt}\, =\, - \frac{1}{r^2}\, 
  G_{\th\th}\, =\, -\, \frac{1}{r^2 \sin^2 \th}\, G_{\vf\vf}.
\label{5.9.2}
\ee

\nit
The conditions (\ref{5.9.1}) are met for any $g(r)$ of the form

\be
g(r)\, =\, a\, +\, \frac{b}{r}\, +\, \frac{c}{r^2}, \hspace{3em}
  h(r)\, =\, \frac{g + r g^{\prime}}{g}.
\label{5.9.3}
\ee

\nit 
It turns out that eqs.(\ref{5.9.3}) indeed provide solutions of the Einstein 
equations satisfying the correct boundary conditions. They are most easily 
obtained from the first Einstein equation (\ref{5.8}): for $hg \neq 0$ it is 
possible to factor out all $h$-dependence and obtain an equation for $g$ only: 

\be
gg^{\prime\prime} + \frac{2}{r} 
  gg^{\prime} - \frac{1}{2}\, g^{\prime\, 2}\, =\, 0. 
\label{5.8.0}
\ee

\nit
There is a unique solution with the required normalization for $r \rightarrow 
\infty$: 

\be
g(r)\, =\, \lh 1 + \frac{m}{2r} \rh^2 ,
\label{5.10}
\ee

\nit
where $m$ is an undetermined constant of integration. Eq.(\ref{5.9.1}) then 
immediately provides us with the solution for $h$:

\be
h(r)\, =\, \frac{2r - m}{2r + m}\, .
\label{5.9}
\ee

\nit
The complete solution for the static, spherically symmetric space-time
in isotropic coordinates therefore is 

\be
ds^2\, =\, - \lh \frac{2r - m}{2r + m} \rh^2\, dt^2\, +\, 
       \lh 1 + \frac{m}{2r}\rh^4\, \lh dr^2 + r^2 d\Og^2\rh.
\label{5.11}
\ee
 
\nit
Clearly this solution matches Minkowski space at $r \rightarrow \infty$ 
by construction, and $h$ and $g$ change monotonically (decreasing and 
increasing, respectively) as a function of $r$ untill $r = m/2$. At that 
value $h = 0$, and the metric has a zero mode. Then the derivation of the 
solution we have presented breaks down, and its continuation to values 
$r < m/2$ requires a careful interpretation, discussed below. In any 
case, it turns out that this singularity of the metric is not physical, as 
invariants of the curvature remain finite and well-defined there. Indeed, 
in the following we construct coordinate systems in which the metric is 
perfectly regular at these points and can be continued to regions inside 
the surface $r = m/2$. In the isotropic coordinate system the apparent 
singularity of the metric for $r = 0$ is another coordinate artifact. It 
disappears if one brings spatial infinity to a finite distance by an 
appropriate coordinate transformation. Details are given in the next 
section.

\section{Discussion \label{S5.3}}

Modulo radial reparametrizations $r \rightarrow \bar{r}(r)$ the line-element 
(\ref{5.11}) is the most general static and spherically symmetric solution
of the source-free Einstein equations matching flat Minkowski space-time at
spatial infinity $(r \rightarrow \infty)$. Therefore in the non-relativistic 
weak-field range $(v \ll 1, r \gg m)$ it should reproduce Newton's law for 
the gravitational field of a spherically symmetric point mass of magnitude 
$M$, according to which the acceleration of a test particle at distance $r$ 
is 

\be
\frac{d^2 \br}{dt^2}\, =\, - \frac{GM}{r^3}\, \br
\label{5.11.1} 
\ee

\nit  
As the spherically symmetric solutions of the Einstein equations 
(\ref{5.11}) are distinguished only by a parameter $m$, in the Newtonian 
limit this parameter must be related to the mass $M$, thereby representing 
the equivalent gravitational mass of the object described by the spherically 
symmetric metric (\ref{5.11}). 

We first show that our solutions reproduce Newtons law in the limit 
$r \gg m$, $v \ll 1$ $(= c)$. The equation of motion for a test particle 
is the geodesic equation (\ref{c2.3}). For the spatial coordinates $x^i$
this becomes 

\be 
\ba{lll}
\dsp{ \frac{d^2 x^i}{d\tau^2} } & = & \dsp{ \lh \frac{d x^0}{d\tau} \rh^2\, 
  \Gm{0}{0}{i}\, +\, \mbox{(terms $\sim O(v/c)$)} } \\
 & & \\
 & \approx & \dsp{ -\frac{h}{g^2}\, \pl_i h\, =\, 
   -\frac{m}{r^3}\, x^i\, \frac{h}{g^3}. }
\ea
\label{5.12}
\ee

\nit
Note that in the limit $v \ll 1$ one has $d x^0 = dt \approx d\tau$, whilst 
for $r \rightarrow \infty$ the factor $h/g^3 \rightarrow 1$. Upon 
the identification $m = GM$ we thus indeed reproduce Newton's equation 
(\ref{5.11.1}). 

Taking into account the dimensions of Newton's constant, as follows from 
(\ref{5.11.1}) and is expressed in eq.(\ref{1.2}), and our natural units
in which $c = 1$, we find that $m$ has the dimension of a length, 
consistent with the observation that $m/r$ should be dimensionless. We 
have already noted that in isotropic coordinates the metric component 
$h$ vanishes at half this distance: $r = r_0 = m/2$. Presently this only 
signals that the isotropic system cannot be extended beyond this range, 
but the surface $r = r_0$ does have a special physical significance: it 
defines the location of the horizon of a spherically symmetric black hole. 
This characteristic value $r_0 = m/2$ of the radial coordinate is known as 
the Schwarzschild radius. \vs

\nit
For the range $m/2 < r < \infty$ the Schwarzschild solution in the form 
(\ref{5.11}) is well-defined, and it matches the asymptotic Minkowski 
metric for large values of $r$. We noted that on the Schwarzschild 
sphere $r = m/2$ the metric is singular, but as a solution of the 
Einstein equations it can be continued to smaller values of $r$ and then 
it is again well-defined in the domain $0 < r < m/2$. However, it turns 
out that this domain of $r$-values does not in any sense represent the 
physical interior of the horizon, as one might expect naively. Rather, 
it easy to establish that the solution (\ref{5.11}) is invariant under 
the transformation

\be
r \rightarrow \frac{m^2}{4r}\, ,
\label{5.13}
\ee

\nit
which maps the domain $(0, m/2)$ to $(m/2, \infty)$. Therefore the 
isotropic coordinate system actually represents a double cover of the 
region between the Schwarzschild radius and the asymptotic region $r 
\rightarrow \infty$. Note that under this transformation $h(r) \rightarrow 
- h(r)$, which can be compensated by an accompanying time-reversal
$t \rightarrow -t$. From these observations it follows that the apparent 
singularity for $r = 0$ disappears as it simply describes another 
asymptotic region like that for $r \rightarrow \infty$ in the original 
coordinate system. 

The existence of such a coordinate system presenting a double cover of the 
space outside the black hole might lead one to suspect, that a test particle 
thrown in radially towards the spherical Schwarzschild surface after a long 
enough time would return to an asymptotic Minkowski region. However, this 
does not happen; in the first place, as a simple calculation to be presented 
later shows, in terms of the time measured by an observer at asymptotic 
infinity it takes an infinite period of time for a test particle starting at 
any finite radial distance $r > m/2$ with any finite velocity even to reach 
the Schwarzschild surface; certainly such an observer will never see it 
start up and set off towards infinity again. Apart from this, leaving the 
Schwarzschild surface would also take an infinite amount of asymptotic 
Minkowski-time. Clearly, these effects are due to the vanishing of $h(r)$ 
on the horizon, which implies that an infinite amount of asymptotic 
Minkowski time $t$ has to pass during any finite period of proper time 
$\tau$. But this is only the minor part of the argument; more important is
the second reason, to wit that time-like geodesics do not flow from values
$r > m/2$ to values $r < m/2$. Instead, as will become clear they flow into a 
new region of space-time described by complex values of $r$.

Some calculational details are given later, but the above arguments at least 
indicate that an observer at spatial infinity is essentially disconnected from 
the domain of space-time beyond the Schwarzschild sphere. Therefore the 
Schwarzschild surface is called a {\em horizon}: from spatial infinity one 
cannot look beyond it. In contrast, things are radically different for an 
observer moving with the test particle towards the horizon. For such an 
observer in free fall, co-moving with the test particle on a radial geodesic 
and starting from an arbitrary point at a finite distance from the 
Schwarzschild surface, only a finite amount of proper time passes until 
the radius $r = m/2$ is reached, although for large radial distances the 
required time grows linearly with the distance. 

Moreover, having reached the horizon the particle (and the observer) can 
continue falling inwards without noticing anything particular, at least 
initially. The only remarkable effect they could discover once they have 
passed beyond the Schwarzschild surface is the impossibility to turn back 
to spatial infinity, no matter how powerful the engines they have at 
their disposal to accelerate and propel them. Therefore their passage to 
the inside of the Schwarzschild sphere represents an irreversible event: 
they are permanently trapped inside the Schwarzschild sphere. For this
reason the Schwarzschild space-time is called a black hole: nothing, not
even radiation, can get out once it has fallen through the horizon, at 
least not in the domain of classical relativity.  

To describe this motion of test particles (i.e.\ the flow of geodesics) 
through the Schwarzschild sphere, one must pass to a different coordinate 
system, one better suited to particles in free fall rather than to the
description of the space-time from the point of view of an asymptotic 
Minkowskian observer at rest at spatial infinity. We make this passage 
in two steps. First we show that in other coordinate systems there 
is a different continuation of space-time beyond the Schwarzschild 
surface, one in which the interior geometry inside the horizon is distinct 
from the outside geometry and can not be mapped back isometrically to 
the exterior. After that we show by explicit calculation of the geodesics, 
that this alternative continuation can be taken to decribe the true 
physical situation for a test particle falling through the Schwarzschild
sphere. 

\section{The interior of the Schwarzschild sphere \label{S5.4}}

To go beyond the region of space-time covered by the isotropic coordinates,
we perform a radial coordinate transformation which instead of making the 
spatial part of the metric conformal to flat space, as in (\ref{5.6}), 
(\ref{5.7}), makes the angular part directly isomorphic to the two-sphere
of radius $\bar{r}$: 

\be
\bar{r}\, =\, r g(r), \hspace{3em} \bar{k}^2(\bar{r}) = \bar{r}^2.
\label{5.14}
\ee

\nit
Then the metric takes the form

\be
ds^2\, =\, - \bar{h}^2(\bar{r})\, dt^2\, +\, \bar{g}^{\,2}(\bar{r})\, 
  d\bar{r}^2\, +\, \bar{r}^2 d\Og^2.
\label{5.15}
\ee

\nit
The explicit form of the radial transformation (\ref{5.14}) gives for the 
coefficients

\be
\bar{h}^2(\bar{r})\, =\, \frac{1}{\bar{g}^{\,2}(\bar{r})}\, =\, 
    1\, -\, \frac{2m}{\bar{r}}.
\label{5.16}
\ee

\nit
This leads to the standard form of the Schwarzschild solution as presented 
in the original papers \ct{Schw,Dros}:

\be
ds^2\, =\, - \lh 1\, -\, \frac{2m}{\bar{r}}\rh\, dt^2\, +\, 
  \frac{1}{\dsp{\lh 1\, -\, \frac{2m}{\bar{r}}\rh}}\, 
  d\bar{r}^2\, +\, \bar{r}^2 d\Og^2.
\label{5.16.1}
\ee

\nit
As to lowest order the transformation (\ref{5.14}) does not include a 
rescaling, we have for large $r$ and $\bar{r}$: 

\be
\frac{1}{\bar{r}}\, =\, \frac{1}{r}\, +\, O\,(\frac{1}{r^2}).
\label{5.17}
\ee

\nit
Therefore in the asymptotic region the Newtonian equation of motion for a 
testparticle in the field of a point mass $M$ still holds in the same form
after replacing $r$ by $\bar{r}$, and in the non-relativistic long-distance
limit the two coordinate systems are indistinguishable. 

Note that the Schwarzschild radius $r = m/2$ is now located at 
$\bar{r} = 2m$. Again the metric components are singular there, but now 
there is a zero mode in the $(tt)$-component and an infinity in the 
$(rr)$-component, in such a way that the determinant of the metric remains 
finite on the horizon. On the other hand, the determinant $g$ does have a 
real zero for $\bar{r} = 0$. This metric singularity was not present in 
the isotropic coordinate system, as it did not cover this region of 
space-time, but it turns out to represent a real physical singularity: the 
curvature becomes infinite there in a coordinate-independent way. 
  
A peculiarity of the metric (\ref{5.15}) is that the components $\bar{h}^2$ 
and $\bar{g}^2$ are not positive definite, but change sign at the 
Schwarzschild radius. This implies that the role of time and radial
distance are interchanged for $\bar{r} < 2m$: $\bar{r}$ is the time 
coordinate and $t$ the radial coordinate. At the same time the metric 
components have become explicitly time dependent (they are functions of 
$\bar{r}$), and therefore the metric is no longer static inside the horizon. 
In return, the metric is now invariant under radial shifts $t \rightarrow t 
+ \del t$. Hence it is the radial momentum $p_t$ which now commutes with the 
Hamiltonian (\ref{5.5.1}), the generator of proper-time translations, and is 
conserved. 

This strange result is the price one has to pay for continuing the 
Schwarzschild geometry to the interior of the Schwarzschild sphere. It is 
quite clear that this could never heve been achieved in the isotropic 
coordinate system (\ref{5.11}), at least not for real values of $r$. 
Together with the appearance of the curvature singularity at $\bar{r} = 0$ 
this shows beyond doubt that the region inside the Schwarzschild sphere 
$\bar{r} < 2m$ has no counter part in the standard isotropic coordinate 
system. 

\vs

\nit
To gain a better understanding of the difference between the continuation 
to $r < m/2$ in isotropic coordinates and $\bar{r} < 2m$ in Schwarzschild
coordinates it is useful to perform yet another radial coordinate 
transformation to a dimensionless coordinate $\rg$ defined by \ct{jw6}

\be
\tanh \frac{\rg}{2}\, =\, h(r), 
\label{5.18.1}
\ee

\nit
with the result that

\be
\bar{r}\, =\, rg(r)\, =\, 2m\, \cosh^2 \frac{\rg}{2}.
\label{5.18}
\ee

\nit
Then the Schwarzschild metric becomes

\be
ds^2\, =\, - \tanh^2 \frac{\rg}{2}\, dt^2\, +\, 
        4m^2\, \cosh^4 \frac{\rg}{2}\, \lh d\rg^2 + d\Og^2 \rh.
\label{5.19}
\ee

\nit
As $\cosh \rg/2 \geq 1$ this expression for the metric, like the isotropic
coordinate system, is valid only outside the horizon $\bar{r} \geq 2m$. 
Moreover, continuing to negative values of $\rg$ we reobtain the double 
cover of the exterior of the Schwarzschild black hole:

\be
\tanh \lh \frac{-\rg}{2} \rh\, =\, -h(r)\, =\, h(\frac{m^2}{4r}).
\label{5.20}
\ee

\nit
Now the interior of the Schwarzschild sphere can also be reparametrized 
in a similar way, taking account of the fact that $h^2$ has become 
negative:

\be
\bar{r}\, =\, m \lh 1 + \cos \sg \rh\, =\, 2m \cos^2 \frac{\sg}{2}, 
  \hspace{3em} \tan^2 \frac{\sg}{2}\, =\, - \bar{h}^2(\bar{r}).
\label{5.21}
\ee

\nit
This parametrization clearly exists only for $0 \leq \bar{r} \leq 2m$. 
In these coordinates the metric takes the form

\be
ds^2\, =\, \tan^2 \frac{\sg}{2}\, dt^2\, +\, 4m^2 \cos^4 \frac{\sg}{2}\,
  \lh - d\sg^2 + d\Og^2 \rh.
\label{5.22}
\ee

\nit
The time-like nature of $\sg$ now is obvious. It is also clear, that 
this coordinate system presents an infinitely repeated {\em periodic}
covering of the interior of the Schwarzschild sphere: the fundamental 
domain may be chosen to be $0 \leq \sg \leq \pi$, corresponding to $2m 
\geq \bar{r} \geq 0$; but continuation of $\sg$ beyond this region leads 
to repeated covering of the same set of $\bar{r}$-values. 

Comparing these two parametrizations of the exterior and the interior of 
the Schwarzschild sphere, it is now obvious that they are related by
analytic continuation \ct{jw6}: the interior of the sphere is not described
by negative values of $\rg$, which correspond to $r < m/2$ in isotropic 
coordinates, but by imaginary values of $\rg$:

\be
\sg\, =\, \pm i \rg,
\label{5.23}
\ee

\nit
which is like a Wick rotation changing the space-like radial coordinate 
$\rg$ to the time-like coordinate $\sg$. In terms of the original isotropic 
coordinates it can be described as the analytic continuation of $r$ to
complex values

\be
r\, =\, \frac{m}{2}\, e^{i \sg}.
\label{5.24}
\ee

\nit
Thus in the interior of the Schwarzschild surface the modulus of $r$ is 
constant, but its phase changes, whilst outside its phase is constant
and the modulus changes. Note that whereas the period of the fundamental
domain of $\sg$ in describing the interior region of the Schwarzschild
sphere is $\pi$, the values of the isotropic radial coordinate $r$ are 
periodic in $\sg$ with period $2\pi$. Again it seems that the isotropic 
coordinates cover the Schwarzschild solution twice. We have more to say 
about that in the following. Finally we observe, that as indicated in 
eq.(\ref{5.23}) there are two ways to perform the analytic continuation 
from $\rg$ to $\sg$, i.e.\ from the exterior to the interior of the 
Schwarzschild space-time. Also this fact has non-trivial significance 
for the geometry of the Schwarzschild space-time. 

\section{Geodesics \label{S5.5}}

The analysis presented so far makes it clear that the geometry of 
the Schwarzschild space-time is quite intricate. Indeed, at the 
Schwarzschild sphere $r = m/2$ (or $\bar{r} = 2m$ and $\sg = \rg =0$, 
respectively) several branches of the space-time meet: two branches 
of the exterior connected to an asymptotic Minkowski region, and two 
branches of the interior connecting the horizon with the space-time 
singularity. It is therefore of importance to understand the flow of 
geodesics in the neighborhood of the horizon, and see which branches 
test particles moving in the Schwarzschild space-time follow. We begin
with time-like geodesics, representing the motion of test particles of
unit mass. For such a particle one has in asymptotic Minkowski space:

\be
2H\, =\, g^{\mu\nu}\, p_{\mu} p_{\nu}\, =\, -1.
\label{5.24.1}
\ee 

\nit
This relation was already argued from the definition of proper time in 
sect.(\ref{S2.1}), eq.(\ref{c2.4}) and below. As $H$ is conserved, the
relation holds everywhere on the particle's worldline. 

The solution of the geodesic equations for Schwarzschild space-time 
is much simplified by the large number of its symmetries. They provide 
us with sufficiently many additional constants of motion to allow a 
complete solution of the equations of motion of test particles in terms 
of their energy and angular momentum. The first of these constants of 
motion is the momentum component $p_t$, conserved because the metric is 
$t$-independent:

\be
p_t\, =\, - h^2(r)\, \frac{dt}{d\tau}\, =\, - \eps \hspace{1em}
  \Leftrightarrow \hspace{1em} \frac{dt}{d\tau}\, =\, \frac{\eps}{h^2(r)}.
\label{5.25}
\ee

\nit
At space-like infinity $h^2 \rightarrow 1$; also, with $\tau$ defined as 
proper time, $dt = d\tau$ for a particle at rest there. Therefore a particle 
at rest at $r \rightarrow \infty$ has $\eps = 1$, equal to the (unit) rest 
mass of the test particle; moreover, as is intuitively obvious and argued 
more precisely below, if it is not at rest it must have $\eps > 1$. Indeed, 
particles with $\eps < 1$ cannot reach spatial infinity and in this sense 
they are in bound states. The factor $\eps$ gives the usual special
relativistic (kinematical) time-dilation for a moving particle in asymptotic
Minkowski space. 

In contrast, for finite radial distance $r_0 < r < \infty$ the factor 
$h^2(r)$ describes the gravitational redshift: the time dilation due 
only to the presence of a gravitational field. It represents a universal 
effect, not influenced by the state of motion of the particle. 

The conservation of angular momentum gives us three conservation laws,
which are however not independent: 

\be
\ba{lll}
J_1 & = & - \sin \vf\, p_\th - \cot \th\, \cos \vf\, p_\vf,  \\
 & & \\
J_2 & = & \cos \vf\, p_\th -\, \cot \th\, \sin \vf\, p_\vf, \\
 & & \\
J_3 & = & p_\vf,
\ea
\label{5.26}
\ee

\nit
with $J_1^2 + J_2^2 + J_3^2 =$ {\boldmath $L$}$^2 \equiv \ell^2$, a 
constant. Here the separate momentum components are

\be
p_\th\, =\, k^2(r)\, \frac{d\th}{d\tau}, \hspace{3em}
p_\vf\, =\, k^2(r) \sin^2 \th\, \frac{d\vf}{d\tau}.
\label{5.27}
\ee

\nit
If one orients the coordinate system such that the angular momentum is in 
the $x^3$-direction, then $\th = \pi/2 =$ constant, and therefore $p_\th 
= 0$ and $p_\vf = J_3 = \ell$. Then the angular coordinates as a function 
of proper time are given by  

\be
\th\, =\, \pi/2, \hspace{3em} \frac{d\vf}{d\tau}\, =\, \frac{\ell}{k^2(r)}.
\label{5.28}
\ee

\nit
It remains to solve for the radial coordinate\footnote{Note that each of 
these equations is still true for arbitrary radial parametrization.} $r$. 
For this we use the conservation of the hamiltonian, setting $2H = -1$:

\be
p_r^2\, =\, \lh g^2\, \frac{dr}{d\tau}\rh^2\, =\,
        g^2\, \lh \frac{\eps^2}{h^2}\, -\, \frac{\ell^2}{k^2}\, -\, 1 \rh.
\label{5.29}
\ee

\nit
Note, that consistency of this equation in the region outside the horizon
requires

\be
\frac{\eps^2}{h^2}\, \geq\, \frac{\ell^2}{k^2}\, +\, 1.
\label{5.29.1}
\ee

\nit
It shows, why in the region $r \rightarrow \infty$ the time dilation factor
has to be greater or equal to unity: $\eps \geq 1$. At the same time, for 
non-zero values of angular momentum $\ell$ the radial distance $r$ can in
general not become arbitrary low: like in the ordinary Kepler problem there 
is a centrifugal barrier to be overcome. For details we refer to the 
literature \ct{MTW,Chan,Nov}. 

Combining eq.(\ref{5.29}) with that for the angle $\vf$, by eliminating 
$\tau$ we directly obtain the orbital equation

\be
\frac{dr}{d\vf}\, =\, \pm \frac{k^2}{\ell g}\, \sqrt{ 
 \dsp{\frac{\eps^2}{h^2}\, -\, \frac{\ell^2}{k^2}\, -\, 1 }}.
\label{5.30}
\ee

\nit
It is clear that an orbit can have a point of closest approach, or furthest 
distance, only if $dr/d\vf = dr/d\tau = 0$.  The only physical solutions 
are points where relation (\ref{5.29.1}) becomes an equality. When the
equality holds identically, for all points of the orbit, the orbit is
circular. One can show that the smallest radius allowed for a stable
circular orbit corresponds in Schwarzschild coordinates to $\bar{r} = 
2 m \ell_0^2 = 6 m$.
\vs

\nit
To the set of time-like trajectories we add the light-like geodesics, 
representing the orbits of massless particles. In that case the proper-time 
interval vanishes identically on any geodesic, that is to say $H = 0$. We 
parametrize the trajectories with an affine parameter $\lb$ and obtain

\be
p_t\, =\, - h^2\, \frac{dt}{d\lb}\, \equiv\, - \gam,
\label{5.30.1}
\ee

\nit
allowing one to replace the affine parameter by $t$ through

\be
d\lb\, =\, \frac{h^2}{\gam}\, dt.
\label{5.30.2}
\ee

\nit
One can again choose the plane of motion to be $\th = \pi/2$, and the 
conservation of angular momentum holds in the form

\be
\frac{\gam r^2}{h^2}\, \frac{d\vf}{dt}\, =\, r^2\, \frac{d\vf}{d\lb}\, 
   =\, \og, 
\label{5.30.3}
\ee

\nit
where $\og$ is a constant. The radial equation now becomes

\be
h^2\, =\, 
  g^2\, \lh \frac{dr}{dt} \rh^2\, +\, k^2\, \lh \frac{d\th}{dt} \rh^2\, 
  +\, k^2 \sin^2 \th\, \lh \frac{d\vf}{dt} \rh^2,
\label{5.31}
\ee

\nit
or using the choice $\th = \pi/2$ above:

\be
g^2\, \lh \frac{dr}{dt} \rh^2\, =\, h^2\, \lh 1 - 
   \frac{\og^2 h^2 k^2}{\gam^2 r^4} \rh.
\label{5.32}
\ee

\nit
Like in the massive case, if $\og \neq 0$ one can eliminate the affine 
parameter $\lb$ also in favour of $\vf$ instead of $t$. 
\vs

\nit
An interesting case, as concerns the exploration of the horizon and 
the connection between the interior and the exterior of the Schwarzschild 
sphere, is that of radial motion: $\ell = 0$ or $\og = 0$, respectively. 
Let us first consider the time-like geodesic for a particle starting from
rest near $r = \infty$, which implies $\eps = 1$ and $dr/dt \leq 0$. Then

\be
\frac{dt}{d\tau}\, =\, \frac{1}{h^2}, \hspace{3em}
\frac{dr}{d\tau}\, =\, - \frac{1}{g h}\, \sqrt{ 1 - h^2 }.
\label{5.33}
\ee

\nit
It is easy to solve these equations in the Schwarzschild coordinates; 
indeed, with $(1 - \bar{h}^2) = 2m/\bar{r}$ one finds a solution valid in 
the whole range $0 < \bar{r} < \infty$, both outside and inside the horizon: 

\be
\frac{\bar{r}(\tau)}{2m}\, =\, \left[ \frac{3}{4m}\, (\tau_0 - \tau) 
      \right]^{\frac{2}{3}},  
\label{5.34}
\ee

\nit
where $\tau_0$ is the proper time at which the particle reaches the 
singularity at $\bar{r} = 0$. Taking this time as fixed, $\tau$ denotes 
the proper time, prior to $\tau_0$, at which the particle is at radial 
distance $\bar{r}$. Observe, that this time is finite for any finite radial 
distance $\bar{r}$; only when starting from infinity it takes an infinite 
proper time to reach the singularity. Nothing special happens at the 
horizon $\bar{r}= 2m$: it is crossed at proper time

\be
\tau_H\, =\, \tau _0\, -\, \frac{4m}{3},
\label{5.35}
\ee

\nit
a proper period $4m/3$ before the singularity is reached (of course, 
in terms of the time-coordinate $\bar{r}$ inside the horizon, it takes a 
finite period $2m$). 

Solving for $t$ as a function of proper time, we have to distinguish two 
cases: $\bar{r} > 2m$ and $\bar{r} < 2m$. Namely, introducing the new 
variable 

\be
y\, =\, \left[ \frac{3}{4m}\, (\tau_0 - \tau) \right]^{\frac{1}{3}},
\label{5.36}
\ee

\nit 
we obtain the differential equation

\be
-\, \frac{1}{4m}\, \frac{dt}{d\tau}\, =\, \frac{y^4}{y^2 -1},
\label{5.37}
\ee

\nit
the solution of which depends on whether the r.h.s.\ is positive or
negative. We find for the two cases the solutions

\be
\frac{t_0 - t}{4m}\, =\, \frac{\tau_0 -\tau}{4m}\, +\, 
   \left[ \frac{3}{4m}\, (\tau_0 - \tau) \right]^{\frac{1}{3}}\, 
   +\, \left\{ \ba{ll} \dsp{
   \arccoth \left[ \frac{3}{4m}\, (\tau_0 - \tau) \right]^{\frac{1}{3}}, }
   & r > 2m; \\ \dsp{
   \arcth \left[ \frac{3}{4m}\, (\tau_0 - \tau) \right]^{\frac{1}{3}}, }
   & r < 2m. \ea \rd
\label{5.38}
\ee

\nit
In both cases the left-hand side becomes infinite at the horizon, i.e.\
at proper time $\tau_H$. This confirms that an external observer at 
infinity will never actually see the particle reach the horizon. 
Similarly, the distance measured from the horizon to the singularity
at $r = 0$ is infinite in terms of the coordinate $t$.
 
Now consider what happens in isotropic coordinates. In terms of a new 
variable 

\be
x\, = \sqrt{\frac{2r}{m}},
\label{5.39}
\ee

\nit
where we choose the positive branch of the square root, integration of 
the equation for $r(\tau)$ is elementary and gives

\be
\frac{2}{m}\, \lh \tau_0 - \tau \rh\, =\, 
         \frac{1}{3}\, \lh x\, +\, \frac{1}{x} \rh^3.
\label{5.40}
\ee

\nit
Clearly this is symmetric under interchange $x \rightarrow 1/x$, 
corresponding to inversion w.r.t.\ the horizon $r = m/2$. Indeed, both
at $r \rightarrow \infty$ and $r = 0$ we see that $\tau \rightarrow 
- \infty$. This confirms our earlier remark that a particle released
from space-like infinity does not return there. Instead, it will 
cross the horizon $x = 1$ at $\tau = \tau_H$, as in the previous 
calculation. Then continuing $r$ to complex values (\ref{5.24}), we
find

\be
x\, =\, e^{\frac{i}{2} \sg}, \hspace{3em}
 \frac{\tau_0 - \tau}{4m} =\, \frac{1}{3}\, \cos^3 \frac{\sg}{2}.
\label{5.41}
\ee

\nit
The singularity $\sg = \pi$ is then reached for $\tau = \tau_0$, as
expected. In terms of the radial coordinate $\rg$, related by the
pseudo-Wick rotation (\ref{5.23}) the equation for the radial distance
in terms of proper time outside the horizon becomes

\be
\frac{\tau_0 - \tau}{4m}\, =\, \frac{1}{3}\, \cosh^3 \frac{\rg}{2}.
\label{5.42}
\ee

\nit
The analytic continuation of coordinates in the complex plane on 
crossing the horizon is manifest. 

Next we consider incoming light-like radial geodesics. The equation is

\be
\frac{dr}{dt}\, =\, - \frac{h}{g}.
\label{5.43}
\ee

\nit
Let $t_0$ refer to the time at which the massless particle (e.g.,
a photon) passes a fixed point on the radius, e.g.\ $\bar{r}_0$ in 
Schwarzschild coordinates. Then the solution of the geodesic equation 
for $t$ is

\be 
\frac{t - t_0}{2m}\, =\, \frac{\bar{r}_0 - \bar{r}}{2m}\, -\, 
       \ln \left| \frac{\bar{r} - 2m}{\bar{r}_0 - 2m} \right|.
\label{5.44}
\ee

\nit
With the absolute value as argument of the logarithm this equation is
valid both outside and inside the horizon. The time for the photon to
reach the horizon from any point outside diverges, as does the distance
it has to move from horizon to the curvature singularity, as measured
by the coordinate $t$. 

\section{Extended Schwarzschild geometry \label{S5.6}}

As we have already noticed that for a massive particle falling across the 
horizon nothing very special happens, it is unsatisfactory that we do not 
have a non-singular description of the same process for massless particles. 
To improve the description of light-like geodesics at the end of the last
section, it is not sufficient to perform a radial coordinate transformation:
one has to eliminate the $t$-coordinate and replace it by one which does not
become infinite on the horizon. Moreover, to describe light-like geodesics 
it would be convenient to have a metric where the light-cone structure is 
manifest by having the radial part of the metric conformal to radial Minkowski
space: a coordinate system in which $(r,t)$ are replaced by coordinates 
$(u,v)$ such that the radial part of the metric becomes 

\be
-h^2(r) dt^2\, +\, g(r)^2 dr^2\, \rightarrow f(u,v) \lh -du^2 + dv^2 \rh.
\label{5.45.0}
\ee

\nit
Such a coordinate system was first proposed by Kruskal and Szekeres \ct{KSz};
defining the new coordinates such as to be dimensionless, the line element
is written in the form 

\be
\frac{ds^2}{4m^2}\, =\, f(u,v)\, \lh -dv^2 + du^2 \rh\, +\, g(u,v)\, d \Og^2,
\label{5.45}
\ee

\nit
with the standard choice for $(u,v)$ defined in terms of the Schwarzschild 
coordinates $(\bar{r},t)$ by:

\be
u^2\, -\, v^2\, =\, \lh \frac{\bar{r}}{2m} - 1 \rh\, 
                    e^{\frac{\bar{r}}{2m}}, 
  \hspace{3em} 
\frac{u}{v}\, =\, \lacc \ba{ll} \dsp{ \tanh \frac{t}{4m},} & 
                                \mbox{if}\, |v| < |u|; \\
                                & \\ 
                                \dsp{ \coth \frac{t}{4m},} &
                                \mbox{if}\, |v| > |u|.  \ea \rd 
\label{5.46}
\ee

\nit
In these coordinates the horizon $\bar{r} = 2m$, $t = \pm \infty$ is 
located at $u = \pm v$, whilst the curvature singularity $\bar{r} = 0$  
corresponds to the {\em two} branches of the hyperbola $u^2 - v^2 = - 1$,
one in the past and one in the future of an observer at space-like infinity. 
More generally, from the first result one concludes that the surfaces
$\bar{r} = constant$ are mapped to hyperbola's $u^2 - v^2 = constant$, 
where the last constant is positive outside and negative inside the 
horizon. These hyperbola's are worldlines of particles in circular orbit, 
i.e.\ being subject to a constant acceleration. 

Similarly hypersurfaces $t = constant$ correspond to $u = constant 
\times v$, represented by straight lines in a $(u,v)$-diagram. Note also 
that the singularity $\bar{r} = 0$ consists of a set of two {\em 
space-like} surfaces in $(u,v)$-coordinates.

More explicitly, we can write $(u,v)$ in terms of $(r,t)$ as

\be
\ba{lll}
u & = & \dsp{ \pm \sqrt{\frac{\bar{r}}{2m} - 1}\, e^{\frac{\bar{r}}{4m}}\, 
              \cosh \frac{t}{4m}, }\\ 
 & & \\
v & = & \dsp{ \pm \sqrt{\frac{\bar{r}}{2m} - 1}\, e^{\frac{\bar{r}}{4m}}\, 
              \sinh \frac{t}{4m}, }
\ea
\label{5.47}
\ee

\nit
outside the horizon $(\bar{r} \geq 2m)$; and 

\be
\ba{lll}
u & = & \dsp{ \pm \sqrt{1 - \frac{\bar{r}}{2m}}\, e^{\frac{\bar{r}}{4m}}\, 
              \sinh \frac{t}{4m}, }\\ 
 & & \\
v & = & \dsp{ \pm \sqrt{ 1 - \frac{\bar{r}}{2m} }\, e^{\frac{\bar{r}}{4m}}\, 
              \cosh \frac{t}{4m}, }
\ea
\label{5.48}
\ee

\nit
within the horizon $(\bar{r} \leq 2m)$. We note once again, that these
coordinates produce two solutions for each region of the space-time 
geometry. In the above conventions the dimensionless coefficient functions 
are given implicitly by 

\be
f(u,v)\, =\, \frac{8m}{\bar{r}}\, e^{- \frac{\bar{r}}{2m}}, 
   \hspace{3em}  g(u,v)\, =\, \frac{\bar{r}^2}{4m^2}.
\label{5.49}
\ee

\nit
We can also compare with the coordinate systems (\ref{5.19}) and 
(\ref{5.22}), describing explicitly the double cover of the 
Schwarzschild space-time:

\be
u\, =\, e^{\frac{1}{2}\, \cosh^2 \frac{\rg}{2}}\, \sinh \frac{\rg}{2}\, 
        \cosh \frac{t}{4m},  \hspace{3em} 
v\, =\, e^{\frac{1}{2}\, \cosh^2 \frac{\rg}{2}}\, \sinh \frac{\rg}{2}\, 
        \sinh \frac{t}{4m},
\label{5.50}
\ee

\nit
for the exterior region $|v| < |u|$; and 

\be
u\, =\, e^{\frac{1}{2}\, \cos^2 \frac{\sg}{2}}\, \sin \frac{\sg}{2}\, 
        \sinh \frac{t}{4m},  \hspace{3em} 
v\, =\, e^{\frac{1}{2}\, \cos^2 \frac{\sg}{2}}\, \sin \frac{\sg}{2}\, 
        \cosh \frac{t}{4m},
\label{5.51}
\ee

\nit
for the interior region $|v| > |u|$. From these expressions we infer, 
that the exterior regions $\rg > 0$ and $\rg < 0$, corresponding to 
$r > m/2$ and $r < m/2$ in isotropic coordinates, and the two interior 
regions $\sg > 0$ and $\sg < 0$ as well, coincide with the regions 
$(u, v) > 0$ and $(u,v) < 0$ in Kruskal-Szekeres coordinates, respectively. 
Hence the double cover of Schwarzschild space-time given by the two 
solutions for the Kruskal-Szekeres coordinates above correspond exactly to 
the double cover we found in isotropic and $(\rg, \sg)$-coordinates. 
 
The Kruskal-Szekeres coordinates were introduced, and are especially useful, 
because they represent the light-like geodesics in a very simple way. In 
particular, the radial light-like geodesic are given by

\be
du^2\, =\, dv^2 \hspace{1em} \Rightarrow \hspace{1em} 
    \frac{du}{dv}\, =\, \pm 1.
\label{5.52}
\ee

\nit
These correspond to straight lines in the $(u,v)$-diagram parallel to
the diagonals. The special light rays $u = \pm v$ are the asymptotes of 
the singularity and describe the horizon itself: they can be interpreted 
as photons permanently hovering on the horizon. Any other light-like 
radial geodesic eventually hits the singularity, either in the past or 
in the future, at finite $(u,v)$, although this is in the region before 
$t = - \infty$ or after $t = \infty$ in terms of the time of an observer 
at spatial infinity. 

As the light cones have a particularly simple representation in the
$(u,v)$-diagram, for all regions inside and outside the horizon,
it follows that time-like and space-like geodesics are distinguished
by having their derivatives $|du/dv| > 1$ for the time-like case,
and $|du/dv| < 1$ for the space-like ones. One can check by 
explicit computation, that a particle starting from rest at any 
distance $r$ and falling into the black hole has a time-like world 
line everywhere, also inside the horizon: $du/dv$ can change on the
world line of a particle, but never from time-like to space-like 
or vice versa. Therefore all time-like geodesics must cross the 
horizon (light-like) and the singularity (space-like) sooner or 
later. If they don't reach or start from spatial infinity, they 
must come from the past singularity across the past horizon, and then 
cross the future horizon to travel towards the final singularity, all
within a finite amount of proper time. 

The Kruskal-Szekeres coordinate system also teaches us more about the 
topology of the Schwarzschild geometry: how the various regions of the
space-time are connected. First of all it shows that there is a
past and a future singularity inside a past and a future horizon 
$(|u| < |v|)$, depending on $v$ being positive or negative. In this 
respect it is helpful to observe, that according to eq.(\ref{5.48}) for 
$\bar{r} < 2m$ the sign of $v$ is fixed for all $t$. In the literature
these regions are sometimes refered to as the {\em black hole}, the interior
of the horizon containting the future singularity, and the {\em white hole}
containing the past singularity. In a standard convention, these regions 
of space-time are classified as regions II (future Schwarzschild sphere) and 
IV (past Schwarzschild sphere), respectively. 

There are also two exterior regions connected to asymptotic Minkowski 
space. They are distinguished similarly by $u > 0$ and $u < 0$, as for
$\bar{r} > 2m$ the sign of $u$ is fixed for all $t$: one cannot reach
negative $u$ starting from positive $u$ or vice versa by any time-like
or light-like geodesic. In the same standard convention these regions are 
refered to as region I and region III, respectively. Being connected by 
space-like geodesics only, the two sheets of the exterior of the Schwarzschild 
space-time are not in causal contact and cannot physically communicate. 
However, the surfaces $t = constant$ do extend from $u = \infty$ to $u = 
- \infty$; indeed, the regions are space-like connected at the locus $u = 
v = 0$, where the two branches of the horizon intersect. To understand the 
topology of Schwarzschild space-time, we must analyze this connection in 
some more detail; this can be done using the technique of embedding surfaces, 
as explained e.g.\ in ref.\ct{MTW}. 

Consider a surface $t = constant$ and fix a plane $\th = \pi/2$. We
wish to understand how the regions $u > 0$ and $u < 0$ are connected
for this plane. The Schwarzschild line element restricted to this plane
is 

\be
ds^2\, =\, \frac{d\bar{r}^2}{ \dsp{ 1 - \frac{2m}{\bar{r}} } }\, +\, 
           \bar{r}^2 d \vf^2.
\label{5.53}
\ee 

\nit
Now we embed this two-dimensional surface in a three-dimensional flat,
euclidean space:

\be
ds^2\, =\, dx^2\, +\, dy^2\, +\, dz^2,
\label{5.54}
\ee

\nit
by taking $x = \bar{r} \cos \vf$, $y = \bar{r} \sin \vf$, and defining the 
surface $z = F(\bar{r})$ such that 

\be
ds^2\, =\, \lh 1 + F^{\prime\, 2}(\bar{r}) \rh\, d\bar{r}^2\, +\, 
\bar{r}^2 d\vf^2.
\label{5.55}
\ee

\nit
Comparison of eqs.\ (\ref{5.55}) and (\ref{5.53}) then implies

\be
\frac{z^2}{16m^2}\, =\ \lh \frac{F(\bar{r})}{4m} \rh^2\, 
                    =\, \frac{\bar{r}}{2m}\, -\, 1.
\label{5.56}
\ee

\nit
This defines a surface of revolution, obtained by taking the parabola
$z^2 = 8mx - 16m^2$ in the $(x,z)$-plane and rotating it around the
$z$-axis. It is a single surface, with looks like a throat, of which the
upper half $z > 0$ corresponds to 

\be
\sqrt{\frac{\bar{r}}{2m} - 1 }\, \geq 0, \hspace{3em} \mbox{or} 
  \hspace{1em} u \geq 0,
\label{5.57}
\ee

\nit
whilst the lower half $z < 0$ describes the branch

\be
\sqrt{\frac{\bar{r}}{2m} - 1 }\, \leq 0,  \hspace{3em} \mbox{or} 
  \hspace{1em} u \leq 0,
\label{5.58}
\ee

\nit
The two regions are connected along the circle $\bar{r} = 2m$ at $z = 0$. 
Of course, the spherical symmetry implies that the same topology 
would be observed for any choice of angle $\th$; any plane with $u > 0$
is connected to one with $u < 0$ in this way. Although no time-like travel 
is possible classically between the regions on each side of the throat, the 
possibility of particles tunneling via quantum processes between the two 
regions poses a curious and interesting question to theories of quantum 
gravity. 

\section{Charged black holes \label{S5.7}}

The spherically symmetric solution of the source-free Einstein equations
can be extended to a solution of the coupled Einstein-Maxwell equations 
including a static spherically symmetric electric field (in the rest
frame of the black hole)\ct{RN}. As the only solution of this type in flat 
Minkowski space is the Coulomb field, we expect a solution of the coupled 
equations to approach the Coulomb potential at space-like infinity:

\be
A\, =\, A_{\mu}dx^{\mu}\, =\, \frac{q}{4\pi r p(r)}\, dt,
\label{5.59}
\ee

\nit
with $p(r) \rightarrow 1$ for $ r \rightarrow \infty$. This is a solution of 
the Maxwell equation 

\be
D_{\mu} F^{\mu\nu}\, =\, 0,
\label{5.59.1}
\ee

\nit
in a spherically symmetric static space-time geometry. The Maxwell field 
strength appearing in this equations is 

\be
F\, =\, \frac{1}{2}\, F_{\mu\nu}\, dx^{\mu} \wig dx^{\nu}\, 
  =\, \frac{q}{4\pi}\, \frac{p + r \pr{p}}{r^2p^2}\, dt \wig dr.
\label{5.60}
\ee

\nit
In isotropic coordinates (\ref{5.7}) the energy-momentum tensor of this 
electro-magnetic field reads

\be
T_{\mu\nu}[F]\, =\, \lh F_{\,\mu\lb} F_{\nu}^{\:\:\lb}\, -\, 
   \frac{1}{4}\, g_{\mu\nu} F^2 \rh\, =\, \frac{q^2}{32 \pi^2}\, 
  \frac{(p + r \pr{p})^2}{r^4 h^2 p^4}\, \mbox{diag} \lh \frac{h^2}{g^2},
  -1,r^2, r^2 \sin^2 \th \rh.
\label{5.61}
\ee
 
\nit
This has the same general structure as the Einstein tensor (\ref{5.9.2}):

\be
G_{\mu\nu}\, =\, \frac{1}{gh}\, \lh 2 h^{\prime} g^{\prime} + 
  \frac{hg^{\prime\, 2}}{g} + \frac{2}{r}\, (hg)^{\prime} \rh\,
  \mbox{diag} \lh \frac{h^2}{g^2}, -1,r^2, r^2 \sin^2 \th \rh,
\label{5.62}
\ee

\nit
which was derived under the conditions (\ref{5.9.1}), with the general
solution (\ref{5.9.3}). Now requiring the coupled Einstein-Maxwell
equations

\be
G_{\mu\nu}\, =\, - 8 \pi G\, T_{\mu\nu}[F],
\label{5.63}
\ee

\nit
this reduces to a single condition on the radial functions $(h,g,p)$,
with the solution

\be
\ba{lll}
p(r) & = & \dsp{ g(r)\, =\, 1\, +\, \frac{m}{r}\, +\, 
                 \frac{\nu^2}{4r^2}, }\\
 & & \\
h(r) & = & \dsp{ \frac{g + r \pr{g}}{g}\, =\, \frac{4r^2 - \nu^2}{
                 4r^2 + 4mr + \nu^2}. }
\ea
\label{5.64}
\ee

\nit
Here the constant $\nu^2$ is defined by 

\be
\nu^2\, =\, m^2 - e^2, \hspace{3em} e^2\, =\, \frac{q^2 G}{4\pi}.
\label{5.65}
\ee

\nit
It may be checked that this solution satisfies the Maxwell equations
in this gravitational fields as well. As $g(r)$ can also be written 
in the form 

\be
g(r)\, =\, \lh 1 + \frac{m}{2r} \rh^2\, -\, \frac{e^2}{4r^2},
\label{5.66}
\ee

\nit
we find that the Schwarzschild solution is reobtained in the limit
$e^2 \rightarrow 0$. For $\nu^2 > 0$ the zero of the metric coefficient 
$h^2$ is now shifted to 

\be
r\, =\, \frac{\nu}{2}.
\label{5.67}
\ee

\nit
As in the Schwarzschild case this is a point where two asymptotically flat 
regions of space-time meet: the metric is again invariant under the 
transformation 

\be
r\, \rightarrow\, \frac{\nu^2}{4r}, \hspace{3em} t\, \rightarrow\, -t,
\label{5.68}
\ee

\nit
under which $h(r)dt \rightarrow h(r)dt$. Moreover, also like in the 
Schwarzschild geometry, the sphere $r = \nu/2$ actually defines a 
horizon and time-like geodesics are continued into the interior of this 
horizon, which is not covered by the (real) isotropic coordinate system. 

A Schwarzschild-like coordinate system which allows continuation into
the interior of the horizon exists. It was found by Reissner and Nordstrom
\ct{RN} and is related to the isotropic coordinates by a radial coordinate 
transformation

\be
\bar{r}\, =\, rg(r)\, =\, r\, +\, m\, +\, \frac{\nu^2}{4r}.
\label{5.69}
\ee

\nit
Then the metric of the charged spherically symmetric black hole takes 
the form

\be
ds^2\, =\, - \bar{h}^2 dt^2\, +\, \bar{g}^2 d\bar{r}^2\, +\, \bar{r}^2 d\Og^2,
\label{5.70}
\ee

\nit
where the coefficients $(\bar{h}, \bar{g})$ are again inversely related:

\be
\bar{h}^2(\bar{r})\, =\, \frac{1}{\bar{g}^2(\bar{r})}\, 
    =\, 1\, -\, \frac{2m}{\bar{r}}\, +\, \frac{e^2}{\bar{r}^2}\,
    =\, \lh 1 - \frac{m}{\bar{r}} \rh^2\, -\, \frac{\nu^2}{\bar{r}^2}.
\label{5.71}
\ee

\nit
The horizon is now seen to be located at $\bar{r} = m + \nu$; in fact, 
there is a second horizon at $\bar{r} = m - \nu$, provided $m > e$. As in 
general $\bar{h}^2$ and $\bar{g}^2$ are not positive definite, at each of 
these horizons the role of radial and time coordinate is interchanged. 
Only for $\nu = 0$, i.e.\ for $m = e$, the situation is different, as 
the two horizons coincide and no interchange of time- and space-like 
coordinates occurs. This special case of Reissner-Nordstrom solution is
refered to in the literature as an {\em extremal} charged black hole. 

The result of the second horizon for non-extremal charged black holes is, 
that the real curvature singularity that is found to exist at $r = 0$ is 
a time-like singularity, for $m > e$. This is expected from the Coulomb 
solution in flat Minkowski space, which contains a time-like singularity 
of the electric field at $r = 0$. Therefore such a black hole looks more 
like a point particle than the Schwarzschild type of solution of the 
Einstein equations.  

The interior region between the horizons can be obtained from the 
isotropic coordinates by complex analytic continuation of $r$ from the
outer horizon by 

\be
r\, =\, \frac{\nu}{2}\, e^{i\sg}.
\label{5.72}
\ee

\nit
The domain of the argument $\sg$ is chosen as $[0,\pi]$. Indeed, for these 
values of $r$ the radial Reissner-Nordstrom coordinate is

\be
\bar{r}\, =\, m\, +\, \nu \cos \sg,
\label{5.73}
\ee

\nit
and takes all values between the two horizons $\bar{r} = m \pm \nu$ precisely 
once. In this parametrization the isotropic Reissner-Nordstrom metric becomes

\be
ds^2\, =\, \frac{\nu^2 \sin^2 \sg}{\lh m + \nu \cos \sg\rh^2}\, dt^2\, 
   +\, \lh m + \nu \cos \sg \rh^2\, \lh - d\sg^2 + d\Og^2 \rh.
\label{5.73.1}
\ee

\nit
Again we observe the role of $\sg$ as the time-like variable, with $t$ 
space-like between the horizons. For $\sg = \pi$ the isotropic coordinate
$r$ becomes negative: 

\be
\sg\, =\, \pi\, \rightarrow\, r\, =\, - \frac{\nu}{2}.
\label{5.74}
\ee

\nit
Thus we discover that the Reissner-Nordstrom space-time can be extended 
consistently to negative values of the radial coordinate. In particular
the curvature singularity at $\bar{r} = 0$ is seen to be mapped to 

\be
r = - \frac{m \pm e}{2}.
\label{5.75}
\ee

\nit
This shows first of all, that there are two (time-like) branches of the 
singularity, and secondly that $r$ can become negative for non-extremal 
black holes with $m > e$. The two branches are found also upon analytic
continuation of $\sg$ in (\ref{5.73.1}) to imaginary values:

\be
\sg \rightarrow \pm i \rg.
\label{5.76}
\ee

\nit
This leads to a parametrization of the Reissner-Nordstrom geometry
for the exterior regions of the form

\be
ds^2\, =\, - \frac{\nu^2 \sinh^2 \rg}{\lh m \pm \nu \cosh \rg \rh^2}\, dt^2
    +\, \lh m \pm \nu \cosh \rg \rh^2\, \lh d\rg^2 + d\Og^2 \rh,
\label{5.77}
\ee

\nit
where the plus sign holds for the exterior region connected to asymptotic
flat Minkowski space, and the minus sign to the region connected to
the curvature singularity. Because of the double-valuedness of the
analytic continuation (\ref{5.76}) there are two copies of each of these
exterior regions involved.

\section{Spinning black holes \label{S5.8}}

The Schwarzschild and Reissner-Nordstrom black holes are static and
spherically symmetric w.r.t.\ an asymptotic Minkowski frame at spatial
infinity. There also exist extensions of these solutions of Einstein
and coupled Einstein-Maxwell equations which are neither static nor
spherically symmetric, but represent rotating black holes with or without 
charge \ct{Ker,New}. These solutions have an axial symmetry around
the axis of rotation, and are stationary: in the exterior region 
the metric is time-independent, and the rate of rotation and all other 
observable properties of the black hole are constant in time. 

The standard choice of coordinates for these metrics, with the axis of rotation
being taken as the $z$-axis and reducing to those of Schwarzschild, or Reissner 
and Nordstrom for the spherically symmetric case of no rotation, is that of 
Boyer and Lindquist \ct{BoyL}. Denoting these coordinates by 
$(t,\bar{r},\th,\vf)$ to emphasize their relation with the spherical coordinate 
systems for the non-rotating metrics, the line element for reads

\be 
\ba{lll}
ds^2 & = & \dsp{\frac{- \brd^2}{\brr^2 + a^2 \cos^2 \th}\, 
    \lh dt - a \sin^2 \th d\vf \rh^2\, +\, \frac{\brr^2 + 
    a^2 \cos^2 \th}{\brd^2}\, \lh d\brr^ 2 + \brd^2 d\th^2 \rh }\\ 
 & & \\
 & & \dsp{ +\, \frac{(\brr^2 + a^2)^2 \sin^2 \th}{\brr^2 + a^2\cos^2\th}\, 
           \lh d\vf - \frac{a}{\brr^2 + a^2} dt \rh^2.  }
\ea
\label{5.78}
\ee

\nit
Here $a$ is a constant parametrizing the deviation of this line element from 
the usual diagonal form of the Schwarzschild-Reissner-Nordstrom metric; its 
physical interpretation is that of the total angular momentum per unit of mass: 

\be
J\, =\, ma.
\label{5.79}
\ee

\nit
As one can always choose coordinates such that the angular momentum points 
along the positive $z$-axis, the angular momentum per unit of mass may be 
taken positive: $a > 0$. For convenience of discussion this will be
assumed in the following. 

The quantity $\brd^2$ in the line element (\ref{5.78}) is short hand for 

\be
\brd^2\, =\, \lh \brr - m \rh^2\, +\, a^2\, +\, e^2\, -\, m^2.
\label{5.80}
\ee

\nit
Note that, like the metric coefficients $\bar{h}^2$ and $\bar{g}^2$ we
have encountered earlier, this quantity is not positive definite everywhere.
Also, the zero's of this function determine the location of the horizons, of 
which there are two (an outer and an inner one) for $m^2 > a^2 + e^2$:

\be
\brr_{\pm\, H}\, =\, m\, \pm\, \sqrt{m^2- a^2 -e^2}.
\label{5.81}
\ee

\nit
The gravitational field represented by the metric (\ref{5.78}) is
accompanied by an electro-magnetic field, the vector potential of
which can be chosen as the one-form

\be
A\, =\, \frac{q\brr}{4\pi \brg^2}\, \lh dt - a \sin^2 \th d\vf \rh.
\label{5.82}
\ee

\nit
For convenience of notation we have here introduced the function $\brg$ defined
as 

\be
\brg^2\, =\, \brr^2\, +\, a^2 \cos^2 \th.
\label{5.83}
\ee

\nit
These fields are solutions of the coupled Einstein-Maxwell equations 
(\ref{5.59.1}) and (\ref{5.63}), with the Maxwell field strength given
by the two form

\be 
\ba{lll}
F & = & \dsp{ - \frac{q}{4\pi \brg^4}\, \lh \brr^2 - a^2 \cos^2 \th \rh\, 
   dr \wig \lh dt - a \sin^2 \th d\vf \rh }\\
 & & \\
 & & \dsp{ +\, \frac{qa}{2\pi}\, \frac{\brr \cos\th
   \sin \th}{\brg^4}\, d\th \wig \lh a dt - (\brr^2 + a^2)d\vf \rh, }
\ea
\label{5.83.1}
\ee

\nit
and with the identification of the quantity $e^2$ as in eq.(\ref{5.65}):

\[
e^2\, =\, \frac{q^2G}{4\pi}.
\label{5.84}
\]   

\nit
To determine the properties of the space-time described by the Kerr-Newman
line element we need to analyse the geodesic flow. Because the spherical 
symmetry of the Schwarzschild and Reissner-Nordstrom solutions is absent, 
and only axial symmetry remains, there is one less independent constant of 
motion from angular momentum: only the component along the axis of rotation
(the $z$-axis, say) is preserved. Fortunately this is compensated by 
the appearance of a new constant of motion, quadratic in momenta, which
replaces the Casimir invariant of total angular momentum \ct{Cart,WP}. 
As a result one can still completely solve the geodesic equations in terms 
of first integrals of motion. We now give some details.  

First we write down the geodesic hamiltonian, effectively representing the
inverse metric $g^{\mu\nu}$:

\be
H\, =\, \frac{1}{2\brg^2}\, \left[ \brd^2 p_r^2\, +\, p_\th^2\, +\, 
    \lh a \sin \th\, p_t + \frac{1}{\sin \th}\, p_\vf \rh^2\, -\, 
    \frac{1}{\brd^2}\, \lh (\brr^2 + a^2)\, p_t + a\, p_\vf \rh^2 \right].
\label{5.85}
\ee

\nit
Clearly, to be well-defined we must assume $\th \neq (0, \pi)$, as the 
$z$-axis represents a coordinate singularity in the Boyer-Lindquist system
where the angle $\vf$ is not well-defined. The above hamiltonian is a constant 
of geodesic flow; recall, that for time-like geodesics it takes the value 
$2 H = -1$, whilst for light-like ones $H = 0$. 

Next there are two Killing vectors, respresenting invariance of the geometry 
under time-shifts and rotations around the $z$-axis:

\be
\ba{lll}
\ve & = & \dsp{ - p_t\, =\, - g_{tt} \frac{dt}{d\lb}\, -\, g_{t\vf} 
 \frac{d\vf}{d\lb}, }\\
 & & \\
\ell & = & \dsp{ p_\vf\, =\, g_{\vf t} \frac{dt}{d\lb}\, +\, g_{\vf\vf} 
 \frac{d\vf}{d\lb}. }\\
\ea
\label{5.86}
\ee

\nit
Of course, for time-like geodesics the affine parameter $\lb$ maybe taken 
to be proper time $\tau$. These equations can be viewed as establishing a 
linear relation between between the constants of motion $(\ve, \ell)$ and 
the velocities $(dt/d\lb, d\vf/d\lb)$:

\be
\lh \ba{c} \ve \\ \ell \ea \rh\, =\, 
          \tG\, \lh \ba{c} \dot{t} \\ \dot{\vf} \ea \rh,
\label{5.87}
\ee

\nit
with the dot denoting a derivative w.r.t.\ the affine parameter $\lb$. 
The determinant of this linear transformation is 

\be
\det \tG\, =\, - g_{tt} g_{\vf\vf}\, +\, g_{t\vf}^2\, 
                 =\,  \brd^2\, \sin^2 \th,
\label{5.88}
\ee

\nit
and changes sign whenever $\brd^2$ does. Therefore the zero's of $\brd^2$ 
define the locus of points where one of the eigenvalues of $\tG$ changes sign, 
from a time-like to a space-like variable or vice-versa; at the same time, 
the coefficient of $p_r^2$ in the hamiltonian (\ref{5.85}) changes sign, 
confirming that these values represent the horizons of the Kerr-Newman 
space-time. It follows from this analysis, that outside the outer horizon, 
in the region connected to space-like infinity, $\det \tG > 0$.

Now if from eqs.(\ref{5.86}) the angular proper velocity $d\vf/d\lb$ is
eliminated, one finds the relation

\be
\ve\, -\, \Og \ell\, =\, \frac{\det \tG}{g_{\vf\vf}}\, \frac{dt}{d\lb},
\label{5.88.1}
\ee

\nit
where the quantity $\Og$ with the dimensions of an angular velocity is
defined by

\be
\ba{lll}
\Og & = & \dsp{ -\frac{g_{t\vf}}{g_{\vf\vf}}\, =\, a\, \frac{\brr^2 + a^2 - 
  \brd^2}{(\brr^2 + a^2)^2 - \brd^2 a^2 \sin^2 \th} }\\
 & & \\
 & = & \dsp{ \frac{a}{\brr^2 + a^2} \cdot \frac{1}{1 + \brd^2 \brg^2\, 
       (\brr^2 + a^2)^{-1}\, (\brr^2 + a^2 - \brd^2)^{-1}}.  }
\ea
\label{5.88.2}
\ee

\nit
Observe, that it is proportional to $a$ with a constant of proportionality 
which for large but finite $\brr$ is positive. Hence $\Og$ represents an 
angular velocity with the same orientation as the angular momentum $J$. 
With a canonical choice of coordinate system this quantity is positive, 
and therefore $\Og > 0$ as well.

It also follows from eq.(\ref{5.88.1}) that for time-like geodesics outside 
the horizon, with $\lb = \tau$ and therefore $dt/d\lb > 0$, an inequality 
holds of the form 

\be
\ve -\Og \ell > 0.
\label{5.88.3}
\ee

\nit
For large $\brr$ near infinity one finds in fact the stronger inequality
$\ve - \Og \ell > 1$. 

Returning to the solution of the geodesic equations, we take into account 
the existence of a constant of motion $K$ such that $2K = K^{\mu\nu} p_\mu 
p_\nu$, with $K^{\mu\nu}$ the contravariant components of a Killing tensor:

\be
D_{\lh \mu \rd} K_{\ld \nu \lb \rh}\, =\, 0.
\label{5.89}
\ee

\nit
The explicit expression for $K$ is 

\be 
\ba{lll}
K & = & \dsp{ \frac{1}{2\brg^2}\, \lh - \brd^2 a^2 \cos^2 \th\, p_r^2\, +\,
   \brr^2 p_\th^2 \rh\, +\, \frac{\brr^2 \sin^2 \th}{2\brg^2}\, 
   \lh a p_t + \frac{p_\vf}{\sin^2 \th} \rh^2 }\\
 & & \\
 & & \dsp{ +\, \frac{a^2 \cos^2 \th}{2 \brg^2 \brd^2}\, 
   \lh (\brr^2 + a^2) p_t + a p_\vf \rh^2. }
\ea
\label{5.90}
\ee

\nit
That $K$ defines a constant of geodesic flow is most easily established by
checking that its Poisson bracket with the hamiltonian $H$ (\ref{5.85}) 
vanishes:

\be
\lacc K, H \racc\, =\, 0.
\label{5.90.1}
\ee

\nit
Using these constants of motion, it is straightforward to establish the
solutions for time-like geodesics in the form

\be
\ba{lll}
\dsp{ \brg^2 \frac{dt}{d\tau} }& = & \dsp{ \ell a\, \lh 1 - 
   \frac{\brr^2 + a^2}{\brd^2}\rh\, +\, \ve \frac{(\brr^2 + a^2)^2\, 
   -\, \brd^2 a^2 \sin^2 \th}{\brd^2}, }\\
 & & \\
\dsp{ \brg^2 \frac{d\vf}{d\tau} }& = & \dsp{ \ell\, \frac{\brd^2 - 
   a^2 \sin^2\th}{\brd^2 \sin^2 \th}\, -\, \ve a\, \lh 1 - 
   \frac{\brr^2 + a^2}{\brd^2}\rh, }\\
 & & \\
\dsp{ \brg^4 \lh \frac{dr}{d\tau} \rh^2 }& = & \dsp{ 
  - \brd^2 \lh 2K + \brr^2 \rh\, +\, \lh \ell a - \ve (\brr^2 + a^2) \rh^2, }\\
 & & \\
\dsp{ \brg^4 \lh \frac{d\th}{d\tau} \rh^2 }& = & \dsp{ 2K - a^2 \cos^2 \th  
  -\, a^2 \sin^2 \th\, \lh \frac{\ell}{a \sin^2 \th} - \ve \rh^2 .}
\ea
\label{5.91}
\ee

\nit
Note that for a particle starting from $r = \infty$ one has $\ve \geq 1$.
These solutions are well-defined everywhere except on the horizons and at 
points where $\brg^2 = 0$. At the latter locus the curvature invariant 
$R_{\mu\nu\rg\sg} R^{\mu\nu\rg\sg}$  becomes infinite and there is a real 
physical singularity. At the horizons both $t$ and $\vf$ change infinitely 
fast, signalling infinite growth of these coordinates w.r.t.\ an observer in
asymptotic flat Minkowski space. 

As with the infinite amount of coordinate time $t$ to reach the horizon 
in the Schwarzschild and Reissner-Nordstrom solutions, this merely 
signifies a coordinate singularity. However, here it is not only $t$ but 
also $\vf$ which becomes infinite. To gain a better understanding of 
the reason for this rotation of geodesics, first notice that according to
the second equation (\ref{5.91}) even for $\ell = 0$ the proper angular 
velocity $d\vf/d\tau$ does not vanish in general, because $\ve \neq 0$. 
Hence there are no purely radial geodesics: any incoming test particle
picks up a rotation. Now computing the angular velocity as measured
by an observer at rest at spatial infinity from the first two equations
(\ref{5.91}) one finds: 

\be
\frac{d\vf}{dt}\, =\, - \frac{\ve g_{t\vf} + \ell g_{tt}}{
   \ve g_{\vf\vf} + \ell g_{t\vf}}.
\label{5.91.1}
\ee

\nit
It follows immediately, that for a geodesic without orbital angular momentum:
$\ell = 0$, the observed angular velocity equals the special reference value 
(\ref{5.88.2}):

\be
\ld \frac{d\vf}{dt} \right|_{\ell = 0}\, =\, \Og.
\label{5.91.2}
\ee

\nit
At the horizon $\brr = \brr_{+H}$, where $\brd^2 = 0$, this becomes

\be
\ld \frac{d\vf}{dt}\right|_{\left[ \mbox{\footnotesize $\ba{l} \ell = 0 \\ 
   \brr = \brr_{+H} \ea $} \right]}\, =\, \Og_H\, =\, 
   \frac{a}{\brr^2_{+H} + a^2}. 
\label{5.91.3}
\ee

\nit 
Comparison with the expression for $\Og$, eq.(\ref{5.88.2}), shows that 
outside the horizon the inequality

\be
\Og\, \leq\, \Og_H,
\label{5.91.4}
\ee

\nit 
holds, with equality only on the horizon. As these quantities are finite, 
the time $t$ can increase near the horizon at an infinite rate only if $\vf$ 
increases at a proportional rate. For this reason both $t$ and $\vf$ have to 
grow without bound as $\brr$ approaches $\brr_{+H}$. 

At space-like infinity the Killing vectors (\ref{5.86}) define time 
translations and plane rotations. However, their nature changes if one 
approaches the horizon. Consider the time translation generated by 
$p_t = -\ve$. The corresponding Killing vector has contravariant and  
covariant components, given by 

\be
\xi_t^{\mu}\, =\, -\del_t^{\mu} \hspace{1em} \Leftrightarrow \hspace{1em}
\xi_{t\mu}\, =\, -g_{t\mu}.
\label{5.92}
\ee

\nit
Here $\del_{\nu}^{\mu}$ is the usual Kronecker delta symbol. Similarly the 
components of the Killing vector for plane rotations, generated by $p_\vf$, 
are 

\be
\xi_\vf^\mu\, =\, \del_\vf^\mu \hspace{1em} \Leftrightarrow \hspace{1em} 
\xi_{\vf\mu}\, =\, g_{\vf\mu}.
\label{5.93}
\ee

\nit
It follows, that these Killing vectors have norms given by

\be
(\xi_t)^2\, =\, g_{tt}\, =\, \frac{- \brd^2 + a^2 \sin^2 \th}{\brg^2},
\label{5.94}
\ee

\nit
and 

\be
(\xi_\vf)^2\, =\, g_{\vf\vf}\, =\, \sin^2 \th\, \frac{(\brr^2 + a^2)^2 - 
  \brd^2 a^2 \sin^2 \th}{\brg^2},
\label{5.95}
\ee

\nit
whilst their inner product is

\be
\xi_t \cdot \xi_\vf\, =\, - g_{t\vf}\, =\, a \sin^2 \th\:  
  \frac{\brr^2 + a^2 - \brd^2}{\brg^2}.
\label{5.95.1}
\ee

\nit 
From these equations several observations follow. First, as the norms and 
inner products of Killing vectors are coordinate independent quantities,
we can write the expressions for the reference angular velocity $\Og$,
eq.(\ref{5.88.2}), and for the observed angular velocity, eq.(\ref{5.91.1}),
in a coordinate independent form:

\be
\Og\, =\, \frac{\xi_t \cdot \xi_\vf}{\xi^2_\vf}, \hspace{3em}
\frac{d\vf}{dt}\, =\, \frac{\xi_t \cdot \xi_{\vf}\, \ve\, -\, 
   \xi_t^2\, \ell}{\xi_{\vf}^2\, \ve\, -\, \xi_t \cdot \xi_{\vf}\, \ell}.
\label{5.95.2}
\ee

\nit
The last equality is equivalent to

\be
\lh \ve - \Og \ell \rh\, \frac{d\vf}{dt}\, 
  =\, \Og \ve\, -\, \ell\, \frac{\xi_t^2}{\xi_\vf^2}\, 
  =\, \Og\, \lh \ve - \Og \ell\, \frac{\xi_t^2}{ \Og^2 \xi_\vf^2}\rh. 
\label{5.95.2.1}
\ee

\nit
Second, with $a > 0$ and excluding the $z$-axis $\th = (0, \pi)$ ---where 
the Boyer-Lindquist coordinates are not well-defined and $\Og$ vanishes--- 
$\Og$ and the invariant inner product $\xi_t \cdot \xi_\vf$ are positive 
outside the horizon. Also the positivity of $\det \tG$ in the exterior region, 
eq.(\ref{5.88}), can be expressed in the invariant form

\be
\Og^2 \geq \frac{\xi_t^2}{\xi_\vf^2} \hspace{2em} \Leftrightarrow \hspace{2em}
  \frac{\xi_t^2}{\Og^2 \xi_\vf^2}\, \leq\, 1,
\label{5.95.3}
\ee

\nit
with the equality holding only on the horizon. Then for $\ell \geq 0$ we find 
from eq.(\ref{5.95.2.1}) that $d\vf/dt \geq \Og$, whilst for $\ell < 0$ the 
angular velocity is less than this reference value: $d\vf/dt < \Og$. As long 
as $\xi_t$ is time-like this is all that can be established; in fact, with 
$\ell < 0$ and $\xi_t^2 < 0$, the absolute value of the angular velocity 
$d\vf/dt$ can become arbitrarily large.

However, it is easily established that the norm of $\xi_t$ changes sign on 
the surface

\be
\brr (\th)\, =\, \brr_0(\th)\, =\, m\, +\, \sqrt{m^2 - e^2 - a^2 \cos^2 \th},
\label{5.96}
\ee

\nit
which, with the exception of the points where the surface cuts the $z$-axis:
$\th = (0, \pi)$, lies {\em outside} the horizon. The region between this 
surface and the horizon is called the ergosphere, and $\xi_t$ is space-like 
there: $\xi_t^2 > 0$. It then follows directly from eqs.(\ref{5.95.2.1}) and 
(\ref{5.95.3}) that in this region the angular velocity is always positive:

\be
\frac{d\vf}{dt}\, >\, 0, \hspace{3em} \mbox{for}\:\: \xi_t^2\, >\, 0.
\label{5.96.1}
\ee

\nit
Thus we have established, that inside the ergosphere any test particle is
dragged along with the rotation of the black hole and no free-falling
observer can be non-rotating there. Related to this, both $\xi_t$ and 
$\xi_\vf$ are space-like Killing vectors there. Nevertheless, outside
the horizon there still exists time-like combinations of these vectors 
even inside the ergosphere: 

\be
\chi\, =\, \xi_t\, -\, \Og \xi_\vf,
\label{5.96.2}
\ee

\nit
which has non-positive norm: $\chi^2 \leq 0$ as a result of the inequality
(\ref{5.95.3}), with equality holding only on the horizon $\brr = \brr_H$. 

\section{The Kerr singularity \label{S5.9}}
 
We continue the analysis of black-hole geometry with a brief discussion of 
the nature of the singularity of spinning black holes. To begin, we 
have already noticed that, like the spherical charged black hole, the 
spinning Kerr-Newman black holes have two horizons, an innner and an 
outer one. As the singularity $\brg^2 = 0$ is located inside the inner
horizon, we expect the singularity to have a time-like character. 

In order to elucidate the topological structure of the space-time
near the singularity, we introduce new coordinates in two steps. 
First we define the Kerr coordinates $(\tV, \tvf)$ as the 
solution of the differential equations

\be
d\tV\, =\, dt\, +\, \frac{(\brr^2 + a^2)}{\brd^2}\, d\brr, \hspace{3em}
d\tvf\, =\, d\vf\, +\, \frac{a}{\brd^2}\, d\brr.
\label{5.97}
\ee

\nit
With these definitions the Kerr-Newman line element becomes 

\be
\ba{lll}
ds^2 & = & \dsp{ - \frac{\brd^2}{\brg^2}\, \lh d\tV -
  a \sin^2 \th d\tvf \rh^2 +\, \frac{\sin^2\th}{\brg^2}\, \left[
  (\brr^2 + a^2) d\tvf - a d\tV \right]^2  }\\
 & & \\
 & & \dsp{ +\, 2 \lh d\tV - a \sin^2 \th\, d\tvf\rh\, d\brr\, +\, 
     \brg^2\, d\th^2. }
\ea
\label{5.98}
\ee

\nit
Now we turn from quasi-spherical coordinates to quasi-Cartesian ones,
known as Kerr-Schild coordinates, by defining

\be
\ba{lll}  
x & = & \lh \brr \cos \tvf - a \sin \tvf \rh\, \sin \th, \\
 & & \\
y & = & \lh \brr \sin \tvf + a \cos \tvf \rh\, \sin \th \\ 
 & & \\
z & = & r \cos \th,
\ea
\label{5.99}
\ee
  
\nit
and we reparametrize the time coordinate $t \rightarrow \tte$ using 

\be
d\tte\, =\, d\tV\, -\, d\brr.
\label{5.100}
\ee

\nit
We then find that at fixed $(\th,\tvf)$ the transformation 

\be  
\ba{lll} 
x & \rightarrow & \pr{x}\, =\, x\, -\, 2 \brr \cos \tvf, \\
 & & \\
y & \rightarrow & \pr{y}\, =\, y\, -\, 2 \brr \sin \tvf, \\ 
 & & \\
z & \rightarrow & \pr{z}\, =\, - z,
\ea
\label{5.101}
\ee

\nit
is equivalent to $\brr \rightarrow \pr{\brr} = - \brr$. This shows, that
like the Reissner-Nordstrom case, the negative values of $\brr$ are
to be taken as part of the physical domain of values. 

Next one introduces the coefficients $k^i$ $(i = 1,2,3)$ by the 
relation

\be
\bk \cdot d \br\, \equiv\, \frac{- \brr\, (x dx + y dy)\, +\, 
    a\, (x dy - y dx)}{\brr^2 + a^2}\, -\, \frac{z dz}{\brr}\, - d\tte,
\label{5.102}
\ee

\nit
and scalar function $u(\br)$:

\be
u(\br)\, =\, \frac{-\brd^2 + \brr^2 + a^2}{\brg^2}\, 
         =\, \frac{2m\brr - e^2}{\brr^2 + \dsp{ \frac{a^2 z^2}{\brr^2}}}.
\label{5.103}
\ee

\nit
With these definitions the metric can be written in the hybrid form 

\be
ds^2\, =\, dx_\mu^2\, +\, u(\br)\, \lh \bk \cdot d\br \rh^2,
\label{5.104}
\ee

\nit 
where the first term on the right-hand side is to be interpreted as in
flat space-time, with $x^{\mu} = (x,y,z,\tte)$. Therefore $u$ and $\bk$
parametrize the deviation from flat space-time. 

Having introduced this new parametrization, the physical singularity
$\brg^2 = 0$, equivalent to $\brr = 0$ and $\th = \pi/2$, is now seen to 
be mapped to the ring 

\be
x^2\, +\, y^2\, =\, a^2, \hspace{3em} z\, =\, 0,
\label{5.105}
\ee

\nit
with the solution

\be
x\, =\, - a \sin \tvf, \hspace{3em} y\, =\,  a \cos \tvf.
\label{5.106}
\ee

\nit
A consequence of this structure is, that in contrast to the Schwarzschild
geometry, geodesic crossing the horizon do not necessarily encounter
the singularity, but can pass it by. 

\section{Black-holes and thermodynamics}

When a test particle falls into the ergosphere of a black hole, the constant 
of motion $\xi_t \cdot p = \ve$ becomes a momentum component, rather than an 
energy. This provides a means of extracting energy from a spinning black hole, 
as first proposed by Penrose \ct{P2}: if a particle enters the ergosphere with 
energy $\ve$, this becomes a momentum; during its stay in the ergosphere the 
momentum can be changed, for example by a decay of the particle into two or 
more fragments, in such a way that the final momentum of one of the fragments 
is larger than the original momentum. If this fragment would again escape,  
which is allowed as the ergosphere is located outside the horizon, it leaves 
with a different energy $\pr{\ve}$ corresponding to its changed momentum. 
This energy is now larger than the energy $\ve$ with which the original  
particle entered. By this process it is possible to extract energy from a 
black hole, even in classical physics. 

It would seem that the Penrose process violates the conservation of energy, 
if there would be no limit to the total amount of energy that can be 
extracted this way. However, it can be established that such a violation 
is not possible and that the process is self-limiting: emission of energy 
is always accompanied by a decrease in the angular momentum (and/or electric
charge) of the black hole in such a way, that after extraction of a finite
amount of energy the black hole becomes spherical. After that there is no 
longer an ergosphere and no energy extraction by the Penrose process is 
possible. 

We will now show how this comes about for the simplest case of an electrically
neutral black hole of the Kerr type. In this case the location of the horizon 
is at

\be 
\brr_H\, =\, m\, +\, \sqrt{m^2 - a^2}\, =\, 
             m\, +\, \frac{1}{m}\, \sqrt{m^4 - J^2}.
\label{5.107}
\ee

\nit
Then the angular momentum of test particles moving tangentially to the 
horizon is

\be
\Og_H\, =\, \frac{a}{\brr_H^2 + a^2}\, =\, \frac{J}{2m} \cdot 
   \frac{1}{m^2 + \sqrt{m^4 - J^2}}.
\label{5.108}
\ee

\nit
This quantity can be interpreted in terms of the area of the spherical 
surface formed by horizon of the black hole:

\be
A_H\, =\, \int \int_{\brr = \brr_H} d\th d\vf \sqrt{g_{\th\th} g_{\vf\vf}}\, 
  =\, 4\pi\, \lh \brr^2_H + a^2 \rh\, =\, \frac{4\pi a}{\Og_H}.
\label{5.109}
\ee

\nit 
Eq.(\ref{5.108}) thus provides an expression for the area of the horizon in 
terms of the mass and the angular momentum of the black hole:

\be
\frac{A_H}{8\pi}\, =\, m^2\, +\, \sqrt{m^4 - J^2}.
\label{5.110}
\ee

\nit
Now we observe that \nl
{\em (i)} the area $A_H$ increases with increasing mass; \nl 
{\em (ii)} at fixed area $A_H$ the mass reaches a minimum for $J = 0$; \nl
{\em (iii)} by absorbing infalling test particles (including Penrose-type 
processes) the black-hole area $A_H$ never decreases. 
 
The first two statements are obvious from eq.(\ref{5.110}). The last one  
follows if we consider the effect of changes in the mass and angular
momentum of the black hole. We assume that the absorption of a test particle 
with energy $\ve$ and angular momentum $\ell$ changes a Kerr black hole 
of total mass $m$ and angular momentum $J$ to another Kerr black hole of 
mass $\pr{m} = m + \ve$ and angular momentum $\pr{J} = J + \ell$. This 
assumption is justified by the uniqueness of the Kerr solution for 
stationary rotating black holes. 

Now under arbitrary changes $(\del m, \del J)$ the corresponding change in 
the area is

\be
\ba{lll}
\dsp{ \frac{\del A_H}{8\pi} }& = & \dsp{ \frac{2m \del m}{\sqrt{m^4 - J^2}}\,
 \lh m^2 + \sqrt{m^4 - J^2} \rh\, -\, \frac{J \del J}{\sqrt{m^4 - J^2}} }\\ 
 & & \\
 & = & \dsp{ \frac{J}{\Og_H \sqrt{m^4 - J^2}}\, \lh \del m - \Og_H \del J \rh.}
\ea
\label{5.111}
\ee

\nit
For an infalling test particle only subject to the gravitational force of
the black hole we have derived the inequality (\ref{5.88.3}). Moreover, on
the horizon such a particle has been shown to have the angular velocity 
$\Og_H$ w.r.t.\ a stationary observer at infinity, independent of the value 
of $\ell$. Hence with $\del m = \ve$ and $\del J = \ell$ we obtain 

\be 
\frac{\del A_H}{8\pi}\, =\, \frac{J}{\Og_H \sqrt{m^4 - J^2}}\, 
  \lh \ve - \Og_H \ell \rh\, > 0.
\label{5.112}
\ee

\nit
Thus we can indeed conclude that by such adiabatic processes, adding 
infinitesimal amounts of energy and angular momentum at a time, the area 
of the black hole never decreases.  

Casting the relation (\ref{5.111}) in the form

\be
\del m\, =\, \frac{\kg}{8\pi}\, \del A_H\, +\, \Og_H \del J,
\label{5.113}
\ee

\nit
where $\kg$ is given by:

\be
\kg\, =\, \frac{4\pi}{mA_H}\, \sqrt{m^4 - J^2},
\label{5.114}
\ee

\nit
it bears a remarkable resemblance to the laws of thermodynamics which states 
that changes of state described by equilibrium thermodynamics are such that 
for isolated systems variations in energy $U$, entropy $S$ and angular 
momentum $J$ are related by 

\be
\del U\, =\, T \del S\, +\, \Og \del J,
\label{5.115}
\ee

\nit
where $T$ is the temperature and $\Og$ the total angular momentum, and moreover 
that in such processes one always has $\del S > 0$. The important step needed 
in a full identification of these relations is to associate the area $A_H$ 
with the entropy and the quantity $\kg$, called the surface gravity, with the 
temperature:

\be
\del S = c \del A_H, \hspace{3em} T = \frac{\kg}{8 \pi c},
\label{5.116}
\ee

\nit
where $c$ is an unknown constant of proportionality. Now if black holes 
have a finite temperature, they should emit thermal radiation. Using a 
semi-classical approximation, it was shown by Hawking \ct{Haw1}, that quantum 
fluctuations near the horizon of a black hole indeed lead to the emission of 
particles with a thermal spectrum. This is true irrespective of the existence 
of an ergosphere, therefore it covers the case of vanishing angular momentum 
as well. The temperature of the radiation Hawking found was 

\be
T_H\, =\, \frac{\kg}{2\pi}.
\label{5.117}
\ee

\nit
This fixes the unknown constant above to the value

\be
c\, =\, \frac{1}{4}.
\label{5.118}
\ee

\nit
It seems therefore, that the thermodynamical description of black holes
is more than a mere analogy. The issue of interpreting the entropy 

\be
S\, =\, \frac{A_H}{4}\, +\, S_0,
\label{5.119}
\ee

\nit
where $S_0$ is an unknown additive constant, in statistical terms by
counting microstates is hotly debated and has recently received a lot of
impetus from developments in superstring theory.   
 
The extension of these results to the case of charged black holes is 
straightforward. The relation between horizon area $A_H$ and angular 
velocity $\Og_H$ remains the as in (\ref{5.109}), but in terms of the 
macroscopic observables $(m,J,e)$ associated with the black hole the
area is now 

\be
\frac{A_H}{8\pi}\, =\, m^2\, -\, \frac{e^2}{2}\, +\, 
      \sqrt{m^4 - J^2 - m^2 e^2}.
\label{5.120}
\ee

\nit
The changes in these observable quantities by absorbing (possibly charged)
test particles are then related by 

\be
\del m\, =\, \frac{\kg}{8\pi}\, \del A_H\, +\, \Og_H \del J\, +\, 
    \Fg_H \del e,
\label{5.121}
\ee

\nit
where the electric potential at the horizon is 

\be 
\Fg_H\, =\, \frac{4 \pi e \brr_H}{A_H},
\label{5.122}
\ee

\nit
and the constant $\kg$ is now defined as

\be
\kg\, =\, \frac{4\pi}{m A_H}\, \sqrt{m^4 - J^2 -m^2e^2}.
\label{5.123}
\ee

\nit
Again, by sending test particles into the black hole the area can 
never decrease: $\del A_H > 0$. In this way the thermodynamical treatment 
is generalized to include processes in which the electric charge changes. 
\vs

\nit
We conclude this discussion with establishing the gravitational 
interpretation of the constant $\kg$, associated with the temperature in 
the thermodynamical description. This quantity is known as the surface
gravity, because in the spherically symmetric case it represents the 
(outward) radial acceleration necessary to keep a test particle hovering
on the horizon without actually falling into the black hole. 

For the case of the Schwarzschild black hole this is easily established.
Define the covariant acceleration as 

\be
\eta^{\mu}\, =\, \frac{Du^{\mu}}{D\tau}.
\label{5.124}
\ee

\nit
As the four velocity of a test particle has negative unit norm: $u^2 = -1$,
it follows that in the four-dimensional sense the acceleration is orthogonal
to the velocity:

\be
\eta \cdot u\, =\, 0.
\label{5.125}
\ee

\nit
Also, a particle at rest in the frame of an asymptotic inertial observer at
infinity, has a velocity four-vector of the form 

\be
u^\mu\, =\, \lh \frac{1}{ \sqrt{-g_{tt}} }, 0, 0, 0 \rh. 
\label{5.126}
\ee

\nit
Combining these two equations and taking into account the spherical symmetry,
a particle at rest must have a purely radial acceleration given by

\be
\eta^{\brr}\, =\, - \frac{1}{2}\, \pl_{\brr} g_{tt}\, =\, \frac{m}{\brr^2}.
\label{5.127}
\ee

\nit
Remarkably, we recover Newtons law! It follows in particular that on the horizon

\be
\eta^{\brr}_H\, =\,  \frac{1}{4m}\, =\, \kg.
\label{5.128}
\ee 

\nit
This quantity is actually a proper scalar invariant, as can be seen from 
writing it as

\be
\kg\, =\, \left[ \eta\, \sqrt{- \xi_t^2}\, \right]_{\brr_H}, 
\label{5.129}
\ee

\nit
with $\eta$ the invariant magnitude of the four acceleration: 

\be
\eta\, =\, \sqrt{g_{\mu\nu} \eta^{\mu} \eta^{\nu}}.
\label{5.130}
\ee

\nit
For the case of the spinning black holes the proper generalization
of this result gives the surface gravity as in eq.(\ref{5.123}).